# SOLPS-ITER numerical evaluation about the effect of drifts in a divertor configuration of ASDEX-Upgrade and a limiter configuration of J-TEXT

H. Wu[1], P. Shi[2], F. Subba[1], H. Sun[2], M. Wischmeier[3], R. Zanino[1], the ASDEX Upgrade Team

1.NEMO Group, Dipartimento Energia, Politecnico di Torino, Corso Duca degli Abruzzi 24, 10129 Torino, Italy

2.United Kingdom Atomic Energy Authority, Culham Centre for Fusion Energy, Culham Science Centre, Abingdon, Oxon OX14 3DB, United Kingdom.

3.Max-Planck-Institut für Plasmaphysik, Boltzmannstraße 2, D-85748 Garching, Germany

## Abstract:

The High Field High Density Region (HFSFD) has been experimentally observed in both divertor and limiter tokamak devices. In order to numerically reproduced the HFSHD region in the limiter tokamak device J-TEXT, we first performed SOLPS-ITER simulations on the J-TEXT limiter tokamak with the activation of drifts, which is associated with the HFSHD region. The validated physical models, which is from ASDEX Upgrade (AUG) divertor modelling including full drifts and currents, were applied. Through a gas puffing rate scan, both attached and detached regimes were numerically obtained in the AUG divertor and J-TEXT limiter. The key plasma parameters in the J-TEXT limiter are evaluated with and without drifts that have a qualitative performance like AUG except the roll-overment of the total ion flux at the targets. The drift effects on the target profiles are investigated in which the maximum electron temperature at the outer targets is 15eV and 5eV respectively. When the outer targets are attached in the J-TEXT limiter and AUG divertor, the drifts result in the partial detachment of the inner targets. In the detached regimes, the drifts decrease the electron temperature on AUG divertor targets. However, for the J-TEXT limiter, the electron temperature only decreases at the far SOL region. The effect of drifts on the neutral density is also presented.

## 1. Introduction

Edge plasma code packages, e.g. SOLPS-ITE [1], UEDGE[2] and SOLEDGE2D[3], are widely used to study boundary plasma behavior in current tokamak devices, e.g. ASDEX Upgrade (AUG)[4] and Alcator C-mod[5], and to the design of future tokamak devices, e.g. DTT [6], SPARC[7] and EU-DEMO[8]. In SOLPS-ITER, two main modules are included: the multi-fluid plasma solver B2[9] for charged species transport toroidal symmetry, and the Monte Carlo code EIRENE[10], which describes kinetic neutral transport.

The High Field Side High Density region, in which the volume electron density $n_e$ is at least 10 times higher than the upstream averaged and target values, has been observed in the Scrape-Off layer(SOL) of the divertor Tokamak device ASDEX Upgrade (AUG)[11], and has been numerically reproduced by SOLPS-ITER with the activation of drifts [4][12]. Recently, a similar high-density front has been observed in the SOL of the limiter tokamak device J-TEXT, in which the high-density front is seven times higher than the density in the low field side region [14].

In order to investigate the formation of high-density front in the J-TEXT, we first performed a SOLPS-ITER numerical simulation on the J-TEXT limiter configuration with the activation of full drifts and currents[15][16]. The effect of drifts in the J-TEXT limiter is evaluated through a preliminary comparison with AUG single-null divertor configuration simulation results, in which the physical models [4] have been validated against experimental data. The rest of the paper is organized as follows. Section 2 describes the modelling setup. Section 3 compares the effect of drifts in the limiter J-TEXT and divertor AUG, not only in the key parameters, e.g., upstream density and temperature etc., but also the targets profiles in the attached and detached regimes. The conclusions are the section 4.

## 2. Modelling setup

The 96×36 quadrilateral mesh for plasma, both for AUG divertor and J-TEXT limiter, and triangular mesh for neutral particle transport are shown in Figure1. It should be mentioned that there is no private flux region (PRF) in J-TEXT limiter, and it only contains the core and SOL computational regions. In this study, the point in where the LCFS is tangentially contacted with the Limiter targets is name as O-point that is similar to the X-point in divertor configuration.

For the AUG divertor, the gas puffing position is at the outer mid-plane and the pumping and settings are inherited from previous study [4]. For the J-TEXT limiter, the gas puffing location is at the bottom of wall according to experiments[14]. Considering the Carbon material of the first wall in the J-TEXT, a pumping albedo with 0.9 is used at all wall elements to mimic the Carbon absorption. Only pure deuterium is considered. At the core boundary, the ion fluxes are equal to the neutral particle flux cross the core boundary and no other core fueling. Other boundary conditions which are not same with our previous modelling[4] are summarized in Table 1. This is because we found with such values, current J-TEXT Limiter simulation results are not far away from experimental measurements, which are helpful for our future validation. The default set of EIRENE[17] reactions in SOLPS-ITER is employed which include Kotov-2008 model [18]and Deuterium neutral-neutral collision [19].

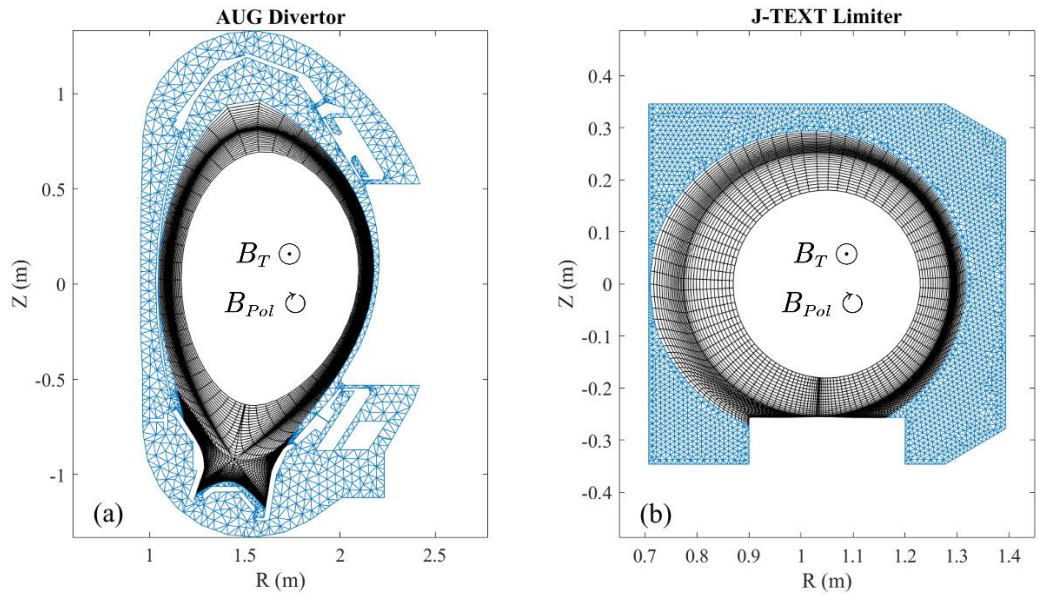

**Figure1 Computational meshes for (a) AUG divertor and (b) J-TEXT limiter.**

Table 1 Boundary conditions for the AUG divertor and J-TEXT limiter simulation.

|  | AUG divertor | J-TEXT limiter |
|---|---|---|
| Input Power | 0.6MW | 0.316MW |
| Leakage factor for ion density at the north boundary | -0.01 | -0.001 |
| Leakage factor for ion temperature at the north boundary | -0.01 | -0.001 |
| Leakage factor for ion thermal velocity at the north boundary | -1.0e-4 | -1.0e-4 |

In this study, the typical AUG L-mode transport coefficients that $D_\perp = 0.5\ m^2/s, \chi_{\perp,ion} = \chi_{\perp,electron} = 1.6\ m^2/s$ [20][21] were selected for both AUG divertor and J-TEXT limiter simulations, instead of adjusting transport coefficients to match experimental upstream profiles. The validated physical models [4][22], including the E×B and diamagnetic drifts and all currents (parallel electric current, anomalous current, diamagnetic current, inertial current, ion-neutral current, current due to perpendicular and parallel viscosity, current due to viscosity tensor), are used for both AUG divertor and J-TEXT limiter.

## 3. Results and discussions

### 3.1 Gas puffing scan

We performed a gas puffing rate scan [23] with and without drifts for both AUG divertor and J-TEXT limiter. The gas puffing rate $\Gamma_{puff,D2}$ is from $1.0\times10^{21}$ D/s to $6.0\times10^{21}$ D/s and from $3.0\times10^{20}$ D/s to $1.2\times10^{21}$ D/s for the AUG divertor and J-TEXT limiter respectively. The performance of key plasma parameters, including outer mid-plane (OMP) electron density $n_{e,omp}$ and OMP temperature $T_{e,OMP}$, maximum electron temperature at the inner and outer targets $T_{e,IT}$ and $T_{e,OT}$, the total ion flux at the inner and outer targets $\Gamma_{ion,IT}$ and $\Gamma_{ion,OT}$, are presented in Figure 2. In Figure 2 (a) and (b) it can be found that as gas puffing rate increase, the $n_{e,omp}$ increase and $T_{e,OMP}$ decrease with and without drifts both in the AUG divertor and J-TEXT limiter. When the gas puffing rate is low, the drifts effect is not strong. However, when the gas puffing value is high, the drifts effect on $n_{e,omp}$ is obvious that result in the $n_{e,OMP}$ is ~15% higher than the non-drift cases. Through gas puffing rate scan with the activation of drifts, the max $T_{e,OT}$ is from 16eV to 3eV and 25eV to 4eV in the AUG divertor and J-TEXT limiter respectively in Figure 2 (c) and (d). From attached regimes to detached regimes [24] are numerically achieved in the AUG divertor and J-TEXT limiter. The differences between with and without drifts are within 20%. From Figure 2 (e) and (f), for the $\Gamma_{ion,IT}$ and $\Gamma_{ion,OT}$, there is no roll-overment in the J-TEXT limier which is only observed in the AUG divertor. This is because the recycled neutrals from J-TEXT limiter targets are directly cross the Last Closed Flux Surface (LCFS) and into the core region, then ionize as ions that transport along the closed magnetic field lines [21]. No recombination front [21][25] is formed in the J-TEXT limiter SOL. Even there is no roll-overment of total ion flux in J-TEXT limiter, with a high gas puffing rate, the max $T_{e,OT}$ can be below 5eV both with and without drifts. Thus, we believe that at such a situation the J-TEXT limiter is in the detach regime [24]. Compared to the non-drifts cases, in the detached regime in both AUG divertor and J-TEXT limiter, the drifts result in a lower total ion flux at both the inner and outer targets.

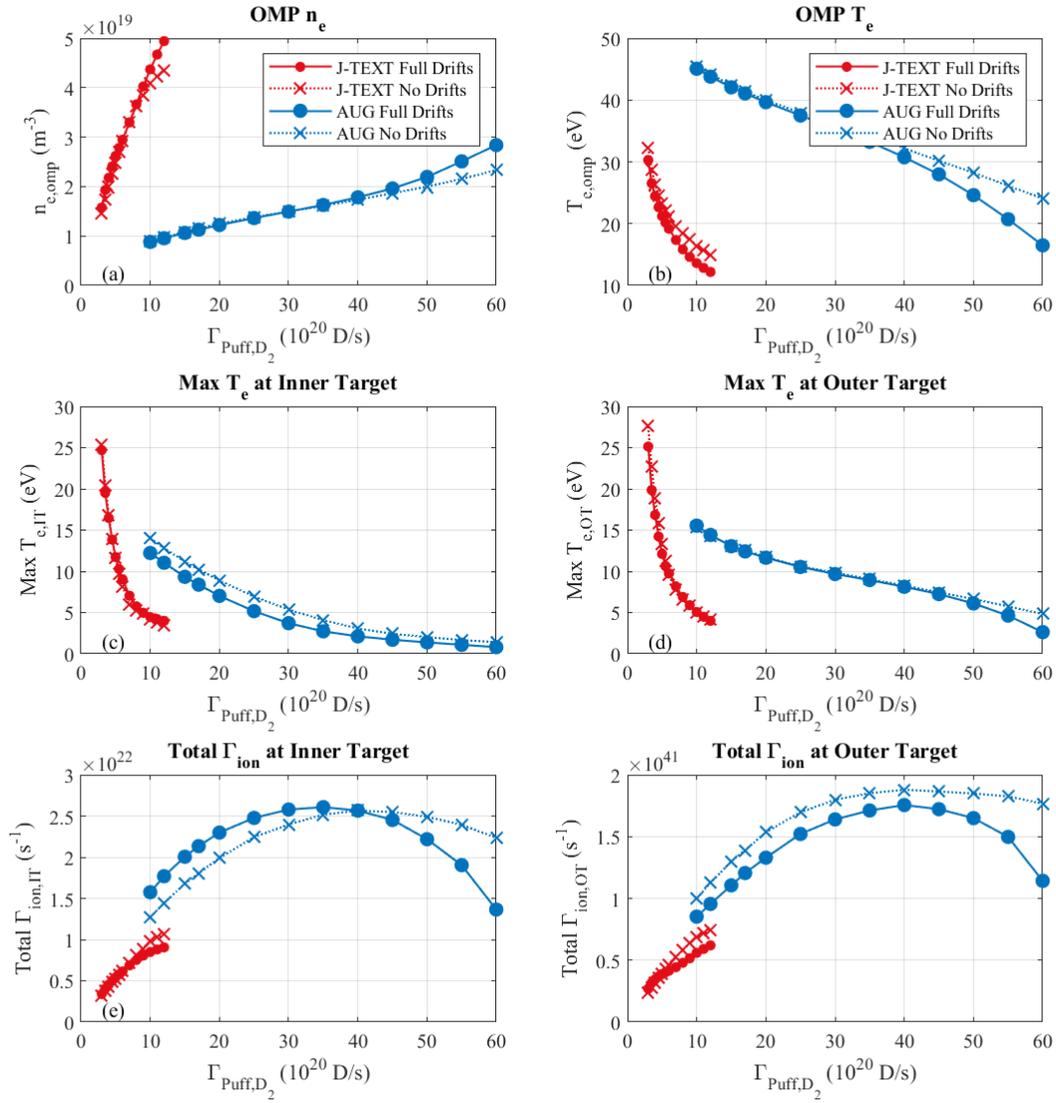

**Figure 2** The AUG divertor and J-TEXT limiter key plasma parameters behavior with and without drifts under $D_2$ gas puffing rate scan, including (a) Electron density at the outer mid-plane $n_{e,OMP}$, (b) Electron temperature at the outer mid-plane $T_{e,OMP}$, (c) Maximum electron temperature at the inner target $T_{e,IT}$, (d) Maximum electron temperature at the at the outer target $T_{e,OT}$, (e) Total ion flux at the inner target $\Gamma_{ion,IT}$ and (f) Total ion flux at the inner target $\Gamma_{ion,OT}$.

## 3.2   Target profiles

In order to evaluate the drifts effect on the targets profiles, two types of cases are considered as attached cases and detached cases, in which the maximum $T_{e,OT}$ (with drifts) are 15eV and 5eV for both AUG divertor and J-TEXT limiter. For the attached cases, the electron pressure balance [26][27] and electron temperature at the targets for the AUG divertor and J-TEXT limiter are shown in Figure 3 and 4. The drifts have no impact on the upstream electron pressure but result in the decrease of electron pressure and temperature at the inner target both for the AUG divertor and J-TEXT limiter, which means when the outer targets are attached, the inner targets are partially detached and this is consistent with previous study[4][12].

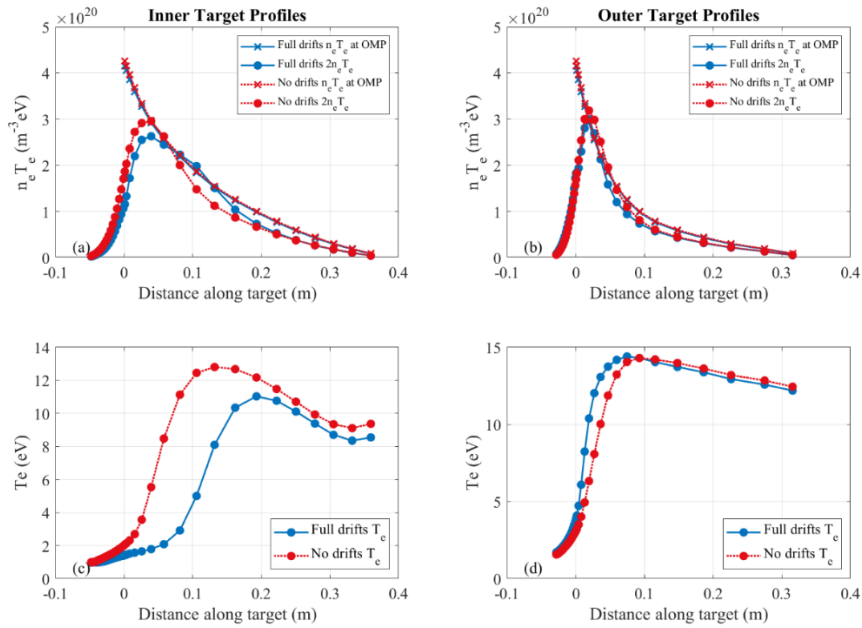

**Figure 3 AUG divertor target profiles for the attached cases. (a) is electron pressure balance at the inner target, (b) is electron pressure balance at the outer target, (c) is the electron temperature at the inner target and (d) is the electron temperature at the outer target.**

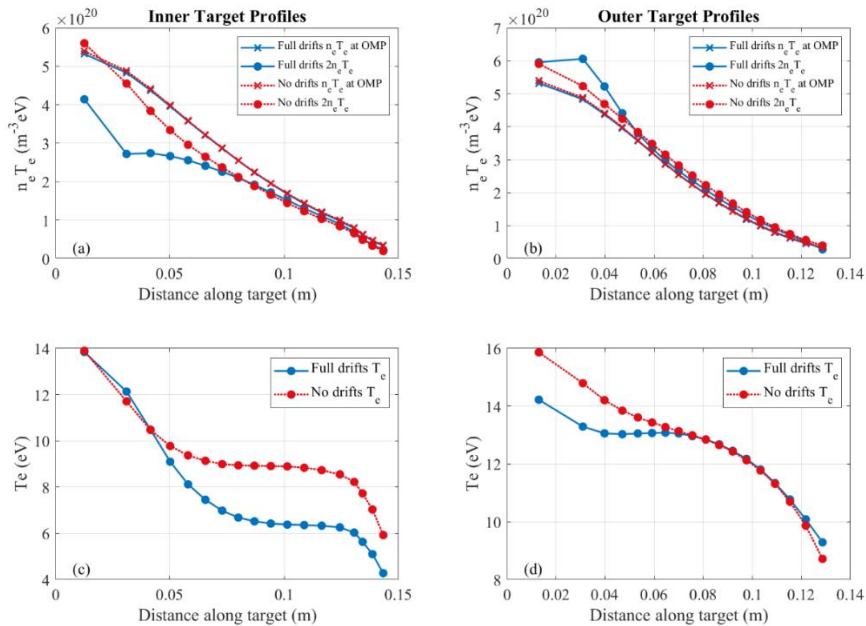

**Figure 4 J-TEXT Limiter target profiles for the attached cases. (a) is electron pressure balance at the inner target, (b) is electron pressure balance at the outer target, (c) is the electron temperature at the inner target and (d) is the electron temperature at the outer target.**

For the detached cases, the simulation results about target profiles are shown in Figure 5 and 6. Compared with non-drift cases, the drifts result in a lower electron pressure at the upstream and targets both in the AUG divertor and J-TEXT limiter. However, the drifts lead to a higher electron temperature near O-point in the J-TEXT limiter, which is the opposite of the AUG divertor. We believe this is because in the detached regime, the recycled neutrals are ionized in the core region that an ionization front is formed in the core region near the O-point. However, with the activation of drifts, the strong poloidal drifts [28] moves the plasma to the upstream. This is also consistent with the fact that drifts increase upstream density in Figure 2. Besides, the length of the cell in the J-JTEXT limiter SOL near O-point is larger than the others that might leads a numeric uncertainty.

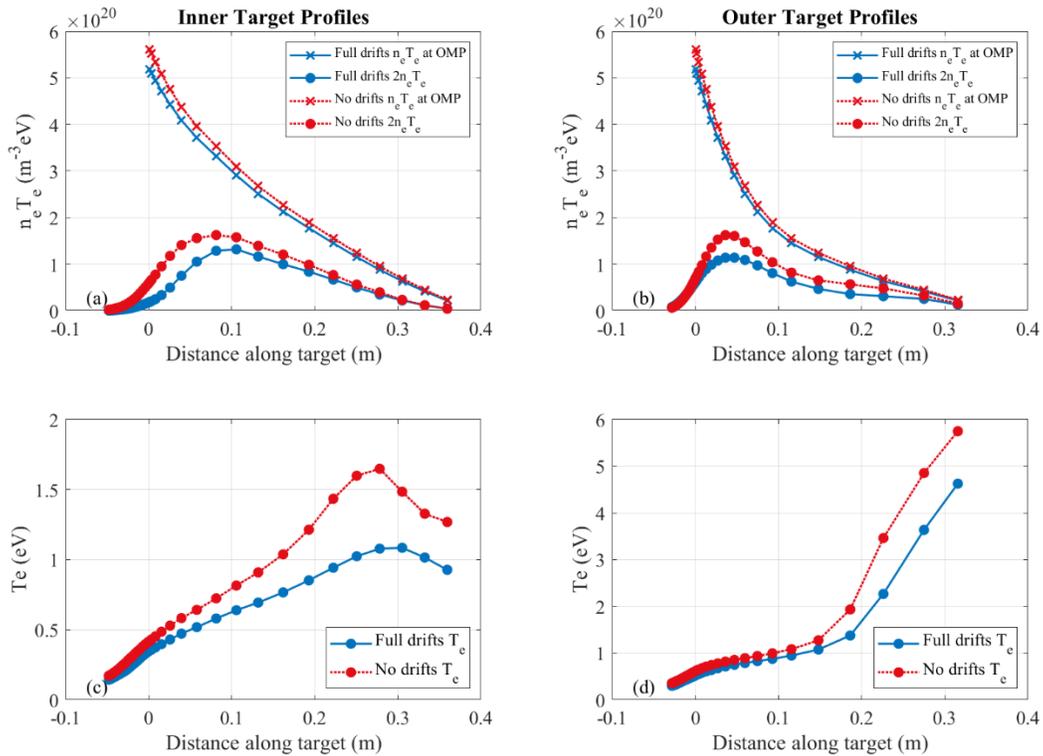

**Figure 5 AUG divertor target profiles for the detached cases. (a) is electron pressure balance at the inner target, (b) is electron pressure balance at the outer target, (c) is the electron temperature at the inner target and (d) is the electron temperature at the outer target.**

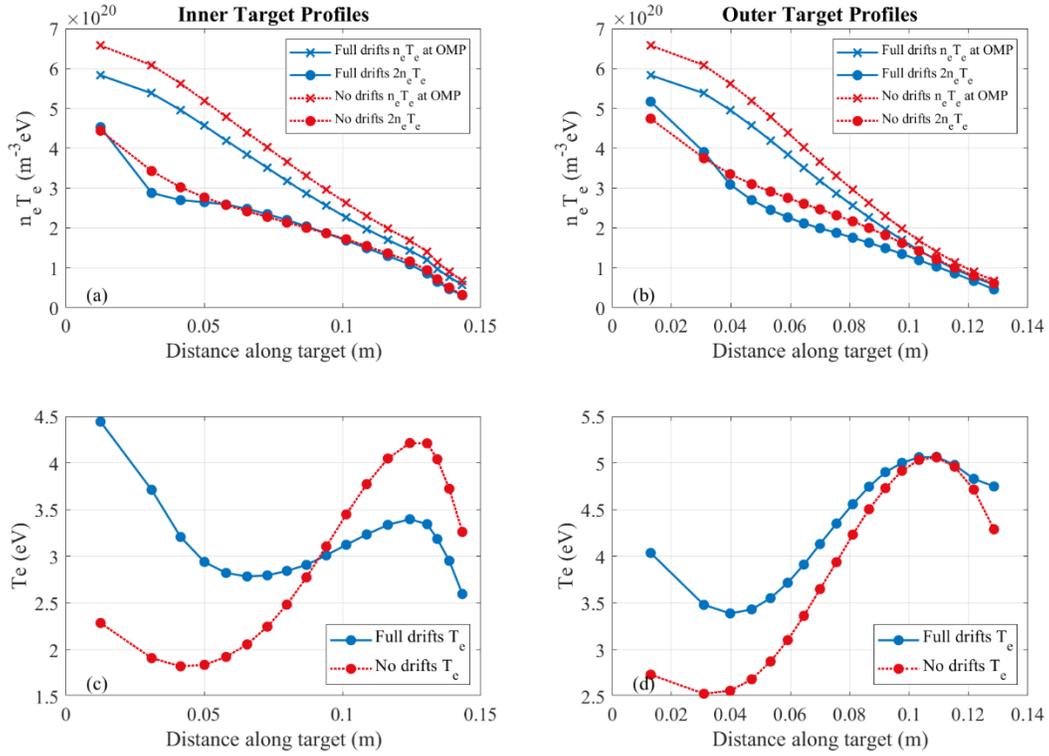

**Figure 6 J-TEXT Limiter target profiles for the detached cases. (a) is electron pressure balance at the inner target, (b) is electron pressure balance at the outer target, (c) is the electron temperature at the inner target and (d) is the electron temperature at the outer target.**

### 3.3 Neutral density distribution

Figure 7 shows the neutral particle density ($D+2D_2$) distributions in the AUG divertor and J-TEXT limiter for the attached and detached cases with the activation of drifts. The neutral density in SOL region near targets in AUG divertor is ~two order magnitude higher than the one in J-TEXT limiter. This is consistent with the fact that the roll-overment of total ion flux is only observed in AUG divertor due to the strong recombination front. The complicated sub-divertor structures in AUG [4][13], which are not exist in the J-TEXT limiter, may also contributes to the high neutral density. Figure 8 shows the ratio of neutral density between with and without drifts. For the attached cases, the drifts result in a higher neutral density in the high field side, which is consistent that the drifts lead to the inner target partially detached. For the detached cases, in AUG divertor, the drifts result in a higher neutral density in the divertor entrance. However, for the J-TEXT limiter, the drifts lead to a lower neutral density near the O-point in the core region and SOL region. In the future, we will investigate in detail about how the ExB drifts and diamagnetic drifts affect the particle flux which is related to the lower neutral density in the J-TEXT limiter.

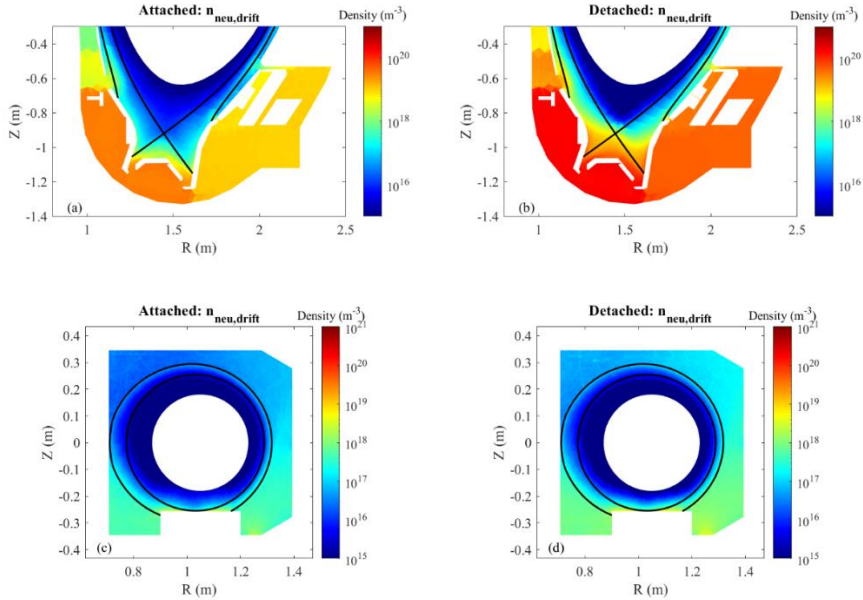

Figure 7 Neutral density (D+2D$_2$) distributions with the activation of drifts. (a) is AUG divertor in attached case, (b) is J-TEXT limiter in attached case, (c) is AUG divertor in detached case and (d) is (c) J-TEXT limiter in detached case.

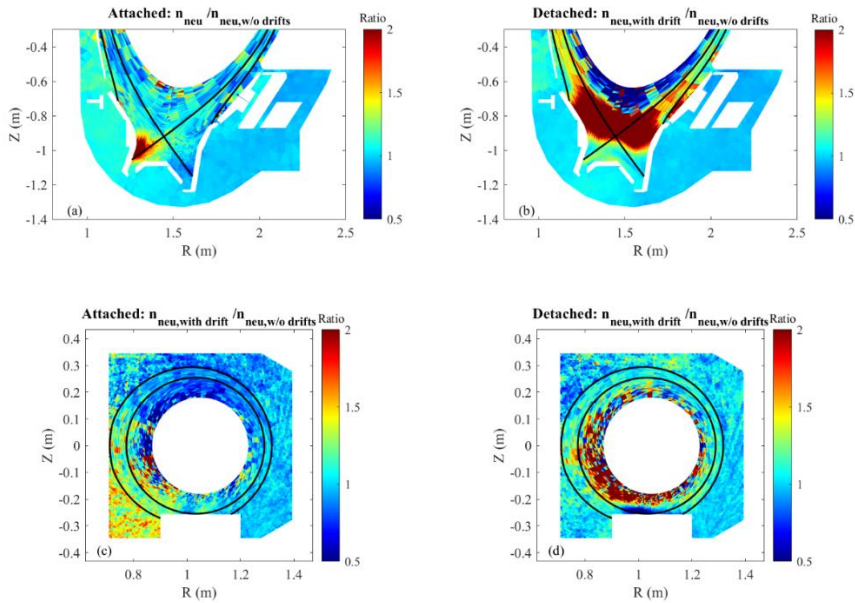

Figure 8 The ratio of neutral density (D+2D$_2$) between with and without drift. (a) is AUG divertor in attached case, (b) is J-TEXT limiter in attached case, (c) is AUG divertor in detached case and (d) is J-TEXT limiter in detached case.

## 4. Conclusions

In this work, SOLPS-ITER numerical simulations on the J-TEXT limiter configuration are performed with and without the activation of drifts. The simulation results are compared with ASDEX upgrade results to assess the effect of drifts. Through a gas puffing rate scan, the key plasma parameters, except the total ion flux, in the J-TEXT limiter have a qualitative performance compared with the AUG divertor simulation results. The recycled neutrals from limiter targets are directly into the core region that there is no recombination front in the limiter SOL. Attached and detached cases are selected, in which the maximum $T_{e,OT}$ are 15eV and 5eV respectively, to evaluate the effect of drifts on the target profiles. In the attached cases, the drifts result in the partial detachment of inner target both in the AUG divertor and J-TEXT limiter. For the detached cases, the drifts decrease electron temperature in the AUG divertor targets but increase the electron temperature near the O-point at the J-TEXT limiter targets. This is because the poloidal drifts move the ions, which is from the ionization of recycled neutrons, to the upstream. The effect of drifts on neutral density distribution is also discussed that the drifts lead to a higher neutral density in the attached regime.

In the future, the J-TEXT limiter simulation results will be validated against experimental data through adjust the transport coefficients. The Carbon impurity and different limiter positions will also be considered.

## Acknowledgements

The authors would like to thank Dr.X. Bonnin for his useful discussions about the neutral particle transport in limiter. This work has been carried out within the framework of the EUROfusion Consortium, funded by the European Union via the Euratom Research and Training Programme (Grant Agreement No 101052200 — EUROfusion). Views and opinions expressed are however those of the author(s) only and do not necessarily reflect those of the European Union or the European Commission. Neither the European Union nor the European Commission can be held responsible for them. This work has been part-funded by the EPSRC Energy Programme (Grant No. EP/W006839/1).